\documentclass[twocolumn]{aastex631}

\usepackage{xcolor}

\usepackage{xspace}
\usepackage{amsmath}

\usepackage{amsmath,amssymb}

\begin{document}
\title{Creating Pileups of Eccentric Planet Pairs Wide of MMRs Through Divergent Migration
}

\author{Jessica Lin}
\affiliation{Department of Physics, Harvey Mudd College, 301 Platt Blvd., Claremont, CA 91711, USA}
\author{Ivan Dudiak}
\affiliation{Department of Physics, Harvey Mudd College, 301 Platt Blvd., Claremont, CA 91711, USA}
\author[0000-0002-1032-0783]{Samuel Hadden}
\affiliation{Canadian Institute for Theoretical Astrophysics, 60 St George Str, Toronto, ON M5S 3H8, Canada}
\author[0000-0002-9908-8705]{Daniel Tamayo}
\affiliation{Department of Physics, Harvey Mudd College, 301 Platt Blvd., Claremont, CA 91711, USA}

\begin{abstract}
Observed pileups of planets with period ratios $\approx 1\%$ wide of strong mean motion resonances (MMRs) pose an important puzzle. Early models showed that they can be created through sustained eccentricity damping driving a slow separation of the orbits, but this picture is inconsistent with elevated eccentricities measured through Transit Timing Variations. We argue that any source of divergent migration (tides, planet-disk interactions etc.) will cause planets that encounter an MMR to both jump over it (piling up wide of resonance) and get a kick to their free eccentricity. We find that the jumps in eccentricity expected from slow MMR crossings are sufficient (but mostly too large) to explain the free eccentricities measured through TTVs. We argue that this mechanism can be brought in line with observations if MMR crossings are not adiabatic and/or through residual eccentricity damping.
\end{abstract}

\section{Introduction}

\begin{figure*}
\centering
\includegraphics[width=0.99\textwidth]{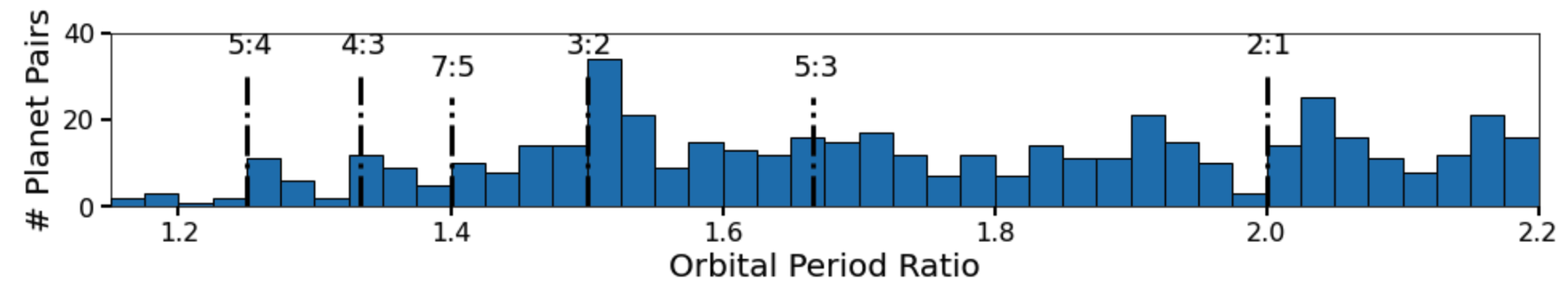}
\caption{The distribution of period ratios between adjacent planets in the NASA Exoplanet Archive. Dash-dotted lines at strong MMRs are labeled with their corresponding integer ratio.}
\label{Kepler-period-ratio}
\end{figure*}

Whether protoplanets grow primarily through the accretion of planetesimals \citep[e.g.,][]{Kokubo00} or through pebble accretion \citep[see review by][]{Johansen17}, planet formation theory typically converges on a final phase of giant impacts and gravitational scatterings that sets the final masses and orbital configurations of the exoplanets we observe today \citep[e.g.,][]{Goldreich04, Izidoro15, Izidoro17, Sanchez24}.
\cite{Tremaine15} proposed that if this phase is sufficiently chaotic, memory of initial conditions should be lost, and the phase space of possible orbital configurations that are dynamically stable should be approximately uniformly filled.
This explains the roughly uniform distribution of period ratios observed between adjacent planet-pairs (Fig.\:\ref{Kepler-period-ratio}, corresponding to effectively random separations between neighboring bodies.

Two exceptions are at the closest separations (period ratios $\lesssim 1.5$), where dynamical instabilities have likely eliminated planet-pairs and significantly carved out the distribution (Chen et al., \textit{submitted}), as well as the pileups of planet-pairs wide of integer ratios corresponding to strong mean motion resonances (MMRs), which were first noted by \cite{Lissauer11}; \citealt{Fabrycky14} and have been a point of research interest. 

An early explanation for these pileups was through an external source of eccentricity damping, which \cite{Lithwick12b} and \cite{Batygin13} showed would naturally cause the orbits of planet pairs in or near MMRs to diverge and move toward period ratios wide of resonance.
While plausible physical mechanisms \citep[e.g., obliquity tides][]{Millholland19} have been identified that could provide the significant eccentricity damping that would be required to create the observed period ratio pileups \citep{Lithwick12b, Batygin13}, such models would also predict that eccentricities would be strongly damped to their equilibrium values.

However, precise measurements of free eccentricities (i.e., deviations from equilibrium eccentricities) extracted from transit-timing variations (TTVs) have revealed values of $\approx 0.01-0.02$ for planet pairs near MMRs \citep{Lithwick12, HL2014,Hadden_2017}, contradicting the eccentricity damping theory \citep{Choksi23, Wu24}.

Other mechanisms (that we return to later in the paper) include the idea that MMR widths grow as planets accrete mass, so that the edge (separatrix) of an MMR could reach a planet pair narrow of resonance and force it to jump across wide of resonance \citep{Petrovich13}, or that non-resonant third planets could be responsible for the non-zero eccentricities \citep{Choksi23}. 

In this paper, we instead investigate the consequences of a simpler picture where the entire population of close-in planets experience divergent migration, rather than just the near-resonant planets. 
This can arise when planets migrate in opposite directions through planetesimal scattering \citep{Chatterjee15, Wu24}, but also when the inner planet migrates inward more than the outer planet, e.g., through tides \citep{Lithwick12} or atmospheric mass loss \citep{Hanf_2024}.

\cite{Lithwick12} argue that such a migration mechanism affecting the whole population can not work, since an initially uniform distribution of period ratios in Fig.\:\ref{Kepler-period-ratio} where all planet pairs moved rightward by a comparable amount will still remain approximately uniform and would create neither troughs nor peaks.

However, as pointed out by the same authors \citep{Wu24}, this argument misses the fact that for a pair of planets migrating further apart from one another (divergently), capture into resonance at integer ratios is impossible \citep[e.g,][]{Murray99}, so the period ratio is forced to jump across the width of the MMR, potentially creating the observed pileups wide of integer ratios.
It could also naturally explain the excited eccentricities of planet pairs in the period ratio pileups, since jumps across MMRs due to divergent migration also naturally provide kicks to the orbital eccentricities \citep{Murray99, Wu24}.

The size of these eccentricity kicks, as well as the widths of MMRs depend on both planets' masses and orbital eccentricities, which complicates quantitative tests across a small number of observations.
In this paper, we use the analytical MMR model of \cite{Hadden19} to non-dimensionalize the eccentricities and period ratio deviations from resonance.
This puts systems with different masses on the same footing, and facilitates both visualization and hypothesis testing.

The paper is organized as follows.
In Sec.\:\ref{sec:TTVs}, we define dynamical variables and introduce our dataset. 
In Sec.\:\ref{normal} we introduce our model and our normalization of observables. 
In Sec.\:\ref{result}, we present our results and discuss mechanisms that could modify the predicted distribution.
Finally, we conclude in Sec.\:\ref{conclusion}.   

\section{Transit Timing Variation (TTV) Data} \label{sec:TTVs}

\subsection{Dynamical Variables}

In our investigation, we use the masses and eccentricities derived from a uniform TTV analysis for 33 pairs of planets analyzed by \citep{Hadden_2017}.
TTV measurements constrain a particular linear combination of the pair of planets' orbital eccentricities called the free eccentricity \citep{Lithwick12}.
For pairs of planets that are not in resonance (the majority), the contribution to the eccentricity that is forced by the resonance is negligible, and the free eccentricity $\mathcal{Z}$ is approximately given by a linear combination of the planets' complex eccentricities $z_i \equiv e_i\mathrm{e}^{i\varpi_i}$, which have a magnitude given by their respective orbital eccentricity $e_i$ and point toward their respective pericenter $\varpi_i$ \citep{Lithwick12}
\begin{equation}
\mathcal{Z} = \frac{f z_1 + g z_2}{\sqrt{f^2 + g^2}}\approx \frac{z_2 - z_1}{\sqrt{2}}. \label{eq:Z}
\end{equation}
The $f$ and $g$ coefficients are specific to each MMR\footnote{In the notation of \citet{hadden_celmech_2022}, $f$ and $g$ are are the disturbing function coefficients $f=C^{(0,0,0,0)}_{(j,1-j,-1,0,0,0)}(\alpha)$ and $g=C^{(0,0,0,0)}_{(j,1-j,0,-1,0,0)}(\alpha)$, evaluated at the resonant semi-major axis ratio $\alpha = \left(\frac{j-1}{j}\right)^{2/3}$. In our calculations, we use the \texttt{celmech} code \citep{hadden_celmech_2022} to evaluate the exact values of $f$ and $g$.}
\footnote{For closely spaced MMRs (period ratio $\lesssim 2:1$, the $f$ and $g$ coefficients are approximately equal and opposite, so the free eccentricity is approximately proportional to the relative eccentricity vector $z_2-z_1$ \citep[for a pedagogical explanation of why this is the physically relevant combination of eccentricity vectors see Sec.\:7.1 of][]{Tamayo_2024}; for the 2:1 MMR $f$ and $g$ coefficients differ significantly from being approximately equal and opposite due to indirect terms \citep[see Table 1 in][]{Deck13}.}.

\begin{figure}[h]
\centering
\includegraphics[width=0.5\textwidth]{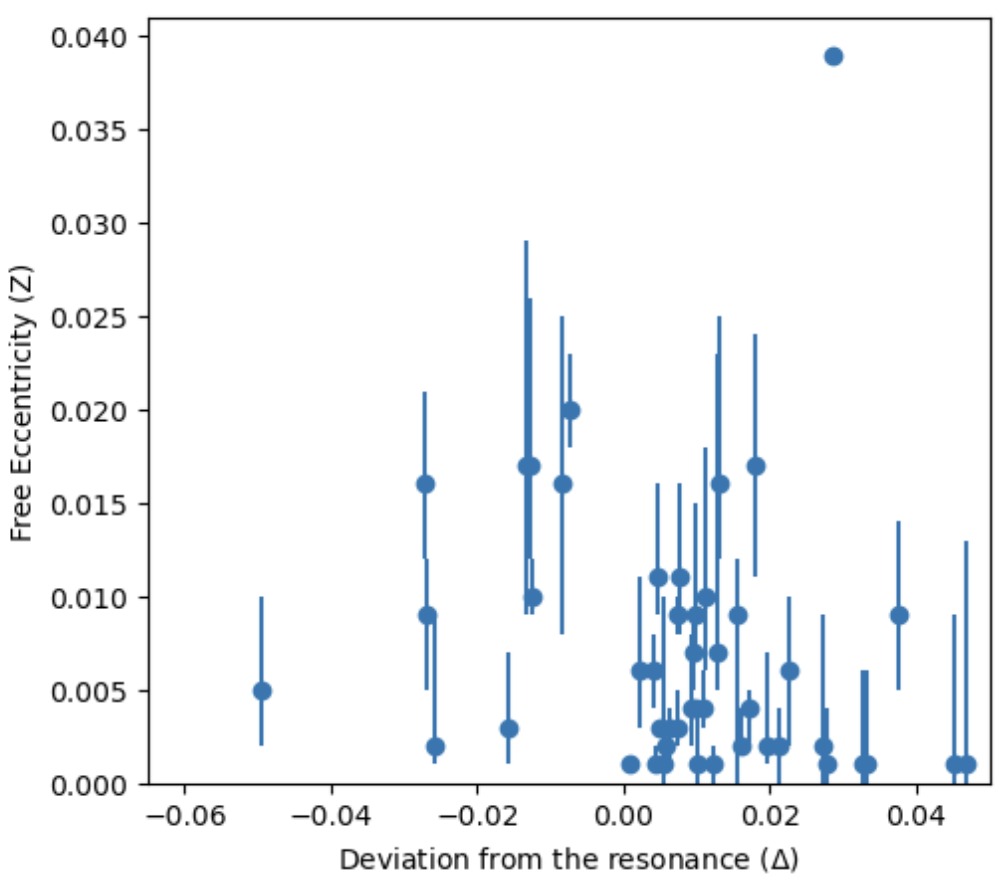}
\caption{Free eccentricities $\mathcal{Z}$ vs. $\Delta$ for planet pairs extracted by \citep{Hadden_2017} (adopting their high-mass priors). \label{fig:data}}
\end{figure}

In Fig.\:\ref{fig:data}, we plot our filtered dataset (see next subsection) of planet pairs' free eccentricities against the fractional deviation of the period ratios from resonance \citep{Lithwick12} 
\begin{equation}
\Delta \equiv \frac{\Bigg(\frac{P_{2}}{P{1}}\Bigg)-\Bigg(\frac{P_2}{P_1}\Bigg)_\text{res}}{\Bigg(\frac{P_2}{P_1}\Bigg)_\text{res}}, \label{eq:Delta}
\end{equation}
where $(P_2/P_1)_\text{res}$ is the integer ratio of the nearest strong MMR. 
This makes it possible to compare planet pairs near different mean motion resonances, and normalizes for the fact that a given period-ratio deviation at very close separations (e.g., the 7:6 MMR) is a larger fractional change than at wider separations (e.g., the 2:1). 
Positive values of $\Delta$ then correspond to planet pairs with period ratios wide of resonance (above the resonant value), while negative values are narrow of resonance.

\subsection{Dataset} \label{sec:dataset}



We analyze the TTV sample from \citep{Hadden_2017}, taking only planet pairs near first-order ($j$:$j-1$) MMRs (within $\Delta = 0.025$). We also remove planet pairs that are inside the resonance itself. Resonant planet pairs are rare in the exoplanet sample \citep[e.g.,][]{Goldreich14}, and likely require a different formation mechanism---presumably via convergent migration \citep[e.g.,][]{Tamayo17}. In particular, we remove Kepler 29 b and c,  Kepler 36 b and c, as well as the resonant chains in Kepler 60, Kepler 80 and Kepler 223. We also remove Kepler 9 b and c, which are outliers at high ($\sim$ Saturn) masses and are very close (or possibly in) a 2:1 MMR \citep{Hadden_2017}.

Finally, we adopt the posteriors from \cite{Hadden_2017} obtained from their high-mass priors, which of their two sets of priors tend to give masses more consistent with observed planetary radii. 


\section{Universal MMR Variables}\label{normal}

One might naively expect eccentricity damping to reduce all eccentricities to zero. 
In fact, eccentricities damp to the non-zero value forced by the MMR, which is itself a function of both how close one is to resonance ($\Delta$) and the planetary masses.
Similarly, the size of the jumps in eccentricity caused by MMR crossings that we investigate in this paper are mass-dependent, making it challenging to evaluate competing predictions in, e.g., Fig.\:\ref{fig:data}, where different planet-pairs have different masses and can not be compared one-to-one.



Additionally, the errors in the masses and eccentricities extracted from TTVs are often strongly correlated \citep{Hadden16}.
Generally, large observed TTVs can be explained both through high masses or high eccentricities, leading to a strong inverse correlation (if the masses are high, the eccentricities are low and vice-versa).
This obfuscates the analysis when projecting the data onto the 2-D space of $|\mathcal{Z}|$ vs. $\Delta$ of Fig.\:\ref{fig:data}.
We therefore seek a way to separate out the dependence on mass (not visible in our plots) in order to compare planet pairs of different masses on the same footing, and avoid confounding effects from strongly correlated parameters.

Because the same resonant dynamics sets both the equilibrium value to which eccentricities should settle under eccentricity damping, as well as the size of the jumps from MMR crossings, we can normalize out this mass-dependence simultaneously to compare both sets of predictions side by side.


\subsection{MMR Jumps}

In this paper, we propose that the free eccentricities of observed planet-pairs near MMRs can be explained through divergent migration ($\dot{\Delta} > 0$), whereby planet pairs encountering an MMR are forced to jump over it and get a kick to their eccentricity.
To check whether sufficiently large eccentricities can be excited through this process, we initialize our planets on circular orbits as a worst-case scenario.

We integrate various such resonance crossings using the Wisdom-Holman \citep{wisdom_symplectic_1991} \texttt{WHFast} integrator \citep{rein_whfast_2015}, which is part of the \texttt{REBOUND} N-body package \citep{rein_rebound_2012}.
To simulate divergent migration, we add exponential semimajor-axis damping to the innermost planet with the \texttt{modify\_orbits\_direct} effect in the REBOUNDx package \citep{tamayo_reboundx_2020}, based on the method of \cite{Lee02}, yielding
\begin{equation}
a_1(t) \approx a_{10}e^{-t/\tau_a},
\end{equation}
where $a_1$ is the inner planet's semimajor axis and $a_{10}$ its initial value. This causes the orbits to separate, so that for $t \ll \tau_a$, $\Delta$ grows at an approximately constant rate
\begin{equation}
\dot{\Delta} \approx \frac{3}{2}\frac{1}{\tau_a}. \label{eq:Deltat}
\end{equation}

\begin{figure*}
\centering
\resizebox{0.99\textwidth}{!}{\includegraphics{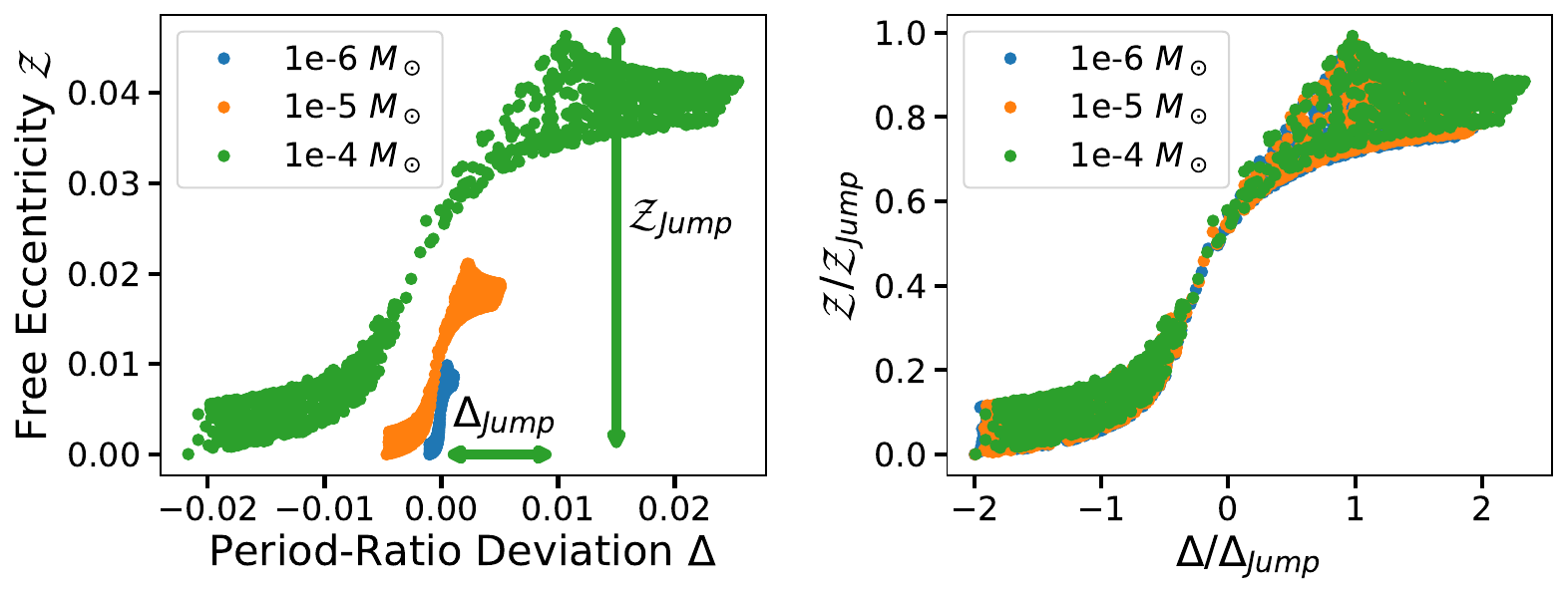}}
\caption{Evolution under divergent migration across the 3:2 MMR for different values of the total planetary mass (different colors) before (left panel) and after (right panel) normalization. 
We plot the free eccentricity ($|\mathcal{Z}|$) vs. the period-ratio deviation from resonance ($\Delta$), with those quantities in the right panel normalized to analytically calculated values (Eqs.\:\ref{eq:zjump} and \ref{eq:deltajump}).} \label{fig:jump}
\end{figure*}

In this picture, the planet pair begins at $\Delta < 0$.
In this regime there is an equilibrium configuration with $\mathcal{Z} \approx 0$, and a second equilibrium at high $\mathcal{Z}$ \citep[e.g.][]{Murray99}.
As $\Delta$ grows, one generically encounters a bifurcation where the equilibrium near $\mathcal{Z} =0$ disappears, and the planet-pair is suddenly forced to oscillate around the high-eccentricity equilibrium.
This causes the planet pair to jump to both a positive $\Delta \equiv \Delta_\textrm{Jump}$, and reach a maximum value $\mathcal{Z}_\textrm{Jump}$.
We illustrate the evolution in the left panel of Fig.\:\ref{fig:jump} for pairs of equal-mass planets crossing the 3:2 MMR around a one solar-mass star.
We ensure that the evolution is adiabatic by setting the migration timescale to be ten times slower than the timescale of the resonant dynamics (we explore the dependence on this migration rate in Sec.\:\ref{rate} below).

Planet pairs with different total mass (colors) undergo different jump magnitudes.
We provide expressions for $\Delta_\textrm{Jump}$ and $\mathcal{Z}_\textrm{Jump}$ as a function of the particular $j$:$j-1$ first-order MMR, as well as the planet masses using the MMR model of \cite{henrard_second_1983, Hadden19}.
Being able to calculate these expressions analytically allows us to normalize the eccentricities and period-ratio deviations from resonance and put systems of different masses on the same footing (right panel of Fig.\:\ref{fig:jump}).
All integrations are initialized with $\Delta = -2 \Delta_\textrm{Jump}$.

\section{Results}\label{result}

The normalized variables introduced above provide a universal prediction for the free eccentricities expected under adiabatic divergent migration through a first-order MMR.
Additionally, because the equilibrium eccentricity $\mathcal{Z}_\textrm{eq}$ expected from strong eccentricity damping is set by the same resonant dynamics with identical mass scaling, we can overplot the predictions of this alternative hypothesis on the same plot, normalizing out all the dependences on the masses and particular resonances.

\begin{figure}[h]
\centering
\includegraphics[width=0.5\textwidth]{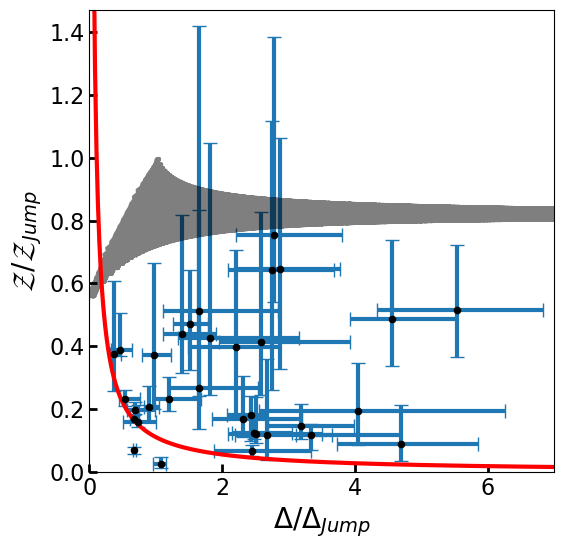}
\label{fig:res_scaled_values}
\caption{Free eccentricities $\mathcal{Z}$ vs. period-ratio deviations from resonance $\Delta$ for observed systems (blue points), with values normalized to those expected from crossing a first-order MMR adiabatically. 
Gray curve shows the full evolution after crossing to $\Delta > 0$, with $\mathcal{Z}/Z_\textrm{Jump} = 1$ denoting the maximum eccentricity reached, and $\Delta/\Delta_\textrm{Jump} = 1$ to the corresponding value of $\Delta$ at the moment of maximum eccentricity.
Migration then continues rightward toward larger values.
In red we plot the equilibrium value of the eccentricity (normalized to $\mathcal{Z}_\textrm{Jump}$). This is the value to which eccentricity damping should drive the observations, and decreases with distance from the resonance.
Most observed planet pairs have eccentricities too high to be consistent with the predictions from strong eccentricity damping.
}
\end{figure}

In Fig.\:\ref{fig:res_scaled_values} we plot these predictions from strong eccentricity damping (red) as well as those from jumps across MMRs (gray), together with our TTV sample from \cite{Hadden_2017} in blue (Sec.\:\ref{sec:dataset}).
To estimate the error bars, we sample 100 random TTV data points from each planet pair's posterior sample given by \cite{Hadden_2017} and plot the 16th to 84th percentile of the normalized values.

We see that observed eccentricities are largely too high to be consistent with the red line expected from the eccentricity damping hypothesis, as also argued previously by \cite{Choksi23, Wu24}.

While the eccentricities are instead too low for our adiabatic predictions for MMR jumps (gray), we point out that no pairs lie above this predicted limit.
We also highlight that the unnormalized free eccentricities are $\sim 0.01$ (see Fig.\:\ref{fig:data}), so in principle they could have been much higher than the gray curve.

We conclude that jumps over MMRs driven by divergent migration provide a plausible mechanism for sufficiently exciting the free eccentricities of near-resonant TTV pairs, though one still needs to account for why the observations lie below the adiabatic predictions in gray.
We consider two possibilities.

\subsection{Non-adiabatic Migration}\label{rate}

The predictions above (described in Appendix A) for the magnitudes of the expected jumps are derived in a Hamiltonian framework where the ``proximity" to resonance $J^\star$ remains approximately constant over the timescale of the resonant dynamics $T_\textrm{res}$ (Eq.\:\ref{eq:Tres}).

When the equilibrium near $\mathcal{Z} = 0$ disappears, $\Delta$ undergoes oscillations of amplitude $\sim \Delta_\textrm{Jump}$(diagonally back and forth in Fig.\:\ref{fig:jump}).
The resonant dynamics therefore drive variations in $\Delta$ at a rate $\dot{\Delta} \sim \Delta_\textrm{Jump}/T_\textrm{res}$.
The adiabatic condition is therefore that the changes in $\Delta$ being driven externally by migration $\dot{\Delta} = \frac{3}{2}\tau_a^{-1}$ (Eq.\:\ref{eq:Deltat}) be long compared to those driven by the resonance.
This requires
\begin{equation}
\tau_a \gg \tau_a^\textrm{crit} \equiv \frac{T_\textrm{res}}{\Delta_\textrm{Jump}}
\end{equation}
where we have dropped factors of order unity.

\begin{figure}[h]
\centering
\includegraphics[width=0.5\textwidth]{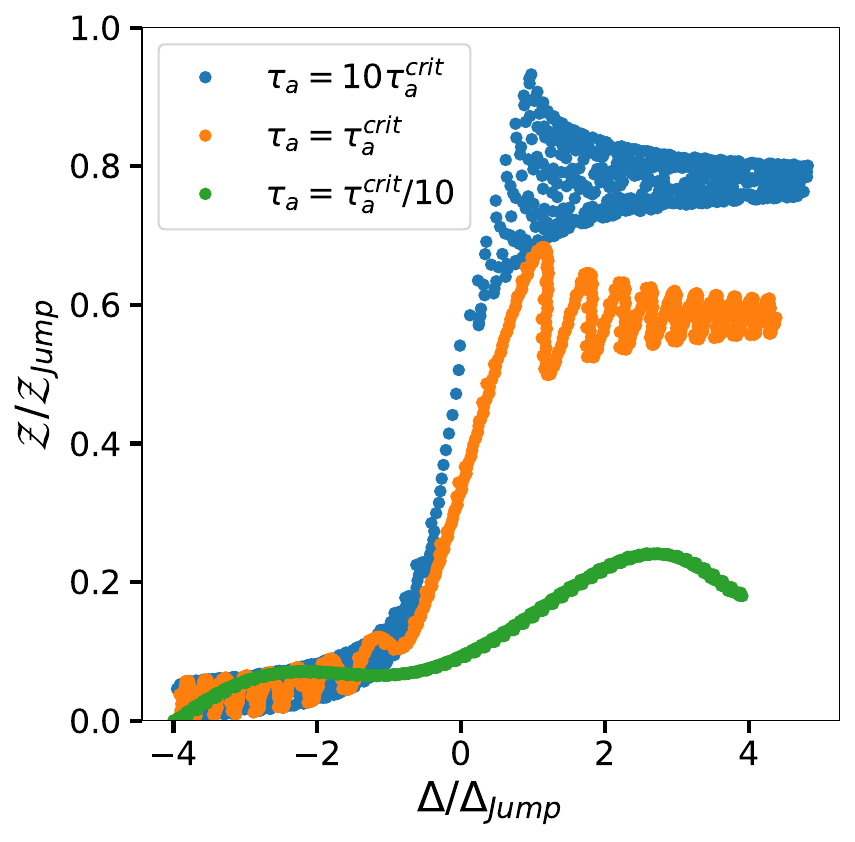}
\caption{Evolution across the 3:2 MMR for two $\approx 1.5 M_\oplus$ planets, varying the migration timescale relative to $\tau_a^\textrm{crit}$ (Eq.\:\ref{eq:tacrit}). 
In the adiabatic limit where $\tau_a \gg \tau_a^\textrm{crit}$ (blue), we recover the full magnitude of the jumps obtained above. For migration timescales comparable to (orange) or shorter (green) than $\tau_a^\textrm{crit}$ the size of the jump is suppressed.
This could help explain the low eccentricities in Fig.\:\ref{fig:res_scaled_values} for observed TTV systems. \label{fig:timescale}}

\end{figure}

We explore this effect in Fig.\:\ref{fig:timescale} for evolution across the 3:2 MMR with two $1.5 M_\oplus$ planets.
If the migration is on a timescale much larger than $\tau_a^\textrm{crit}$ (blue), we recover the adiabatic limit explored above, where $\mathcal{Z}$ approximately reaches $\mathcal{Z}_\textrm{Jump}$.
However, for shorter migration timescales, the size of the jump is reduced. 
In this picture, the adiabatic limit thus provides a theoretical maximum beyond which no observed systems should be found (gray curve in Fig.\:\ref{fig:res_scaled_values}), but faster, non-adiabatic migration can explain the suppressed values of the free eccentricities.

To get a sense for the timescales, we consider the 3:2 MMR as a representative example. 
The resonant scalings depend mostly on the sum of the planetary masses and not on how the mass is distributed between them \citep[e.g.,][]{Deck13}.
Using the expressions in Appendix A, we find
\begin{equation}
\tau_a^\textrm{crit} \approx 3 \times 10^5 \Bigg(\frac{M_{tot}}{2M_\oplus}\Bigg)^{-4/3} \Bigg(\frac{M_\star}{M_\odot}\Bigg)^{4/3} P_{orb}, \label{eq:tacrit}
\end{equation}
where $M_{tot}$ is the total mass of the two planets, $M_\star$ the stellar mass, and $P_{orb}$ the inner planet's orbital period.

Many migration mechanisms, e.g., Type I migration \citep[see, e.g., review by][]{Kley12}, are likely faster than this timescale, so it is plausible that such MMR crossings would not be adiabatic and could help explain the suppressed eccentricities seen in Fig.\:\ref{fig:res_scaled_values}.

\subsection{Partial Eccentricity Damping} \label{sec:edamping}

\cite{Lithwick12b} and \cite{Batygin13} show that eccentricity damping alone can drive migration on much longer timescales.
These authors therefore find that creating the pileups at period ratios $\approx 1\%$ wide of resonance can require $\gtrsim 100$ eccentricity-damping timescales.

In contrast to this picture, many migration mechanisms, e.g., planet-disk interactions \citep{Kley12}, drive both changes in the semimajor axes and the eccentricities.
In the context of Type I migration, the ratio of the timescales for semimajor axis ($\tau_a$) and eccentricity ($\tau_e$) evolution depends (among other variables) on the disk's aspect ratio \citep[see, e.g., Eqs. 15 and 16 of][]{Goldreich14}.
While the semimajor axis generically evolves more slowly than the eccentricity, the ratio $K=\tau_a/\tau_e$ of these two timescales can vary widely.
Many authors adopt a nominal ratio of $K= 100$ \citep{Lee02}. 

In this picture, a $\sim 1\%$ shift in the period ratios (and thus in the semimajor axes) would then correspond to a hundred-fold larger, i.e., order-unity, e-folding decay of the eccentricities.
We show an illustrative example of an N-body integration adopting $\tau_e = \tau_a/100$.

This scenario therefore still corresponds to one of divergent migration across first-order MMRs and obtaining a kick to the free eccentricity.
It simply considers the fact that most migration mechanisms will also introduce some eccentricity damping.
This is in contrast to a picture where eccentricity damping is itself driving a much slower migration, requiring $\sim 100$ eccentricity-damping timescales that would damp the free eccentricities to the red curve in Fig.\:\ref{fig:res_scaled_values}.

\begin{figure}[h]
\centering
\includegraphics[width=0.5\textwidth]{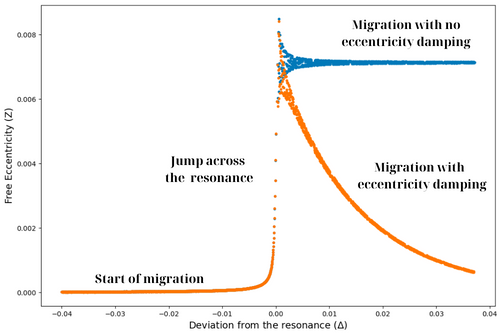}
\caption{N-body integration crossing the 3:2 MMR with an eccentricity damping timescale $\tau_e = tau_a/100$.} \label{fig:edamping}
\end{figure}

We show an illustrative example in Fig.\:\ref{fig:edamping}.

\section{Conclusion}\label{conclusion}

While the observed pileups of planet pairs $\approx 1\%$ wide of strong MMRs can be created through sustained eccentricity damping driving residual migration \citep{Lithwick12b, Batygin13}, this hypothesis seems inconsistent with the elevated free eccentricities observed relative to the equilibrium values predicted in red in Fig.\:\ref{fig:res_scaled_values} \citep[see also][]{Choksi23, Wu24}.

We propose that other mechanisms for divergent migration (e.g., tides, disk-planet interactions) cause planet pairs to encounter and jump over strong MMRs, creating pileups wide of MMR, and simultaneously exciting their free eccentricities.
We show that the jumps in eccentricity expected from adiabatic crossings of such resonances (gray line in Fig.\:\ref{fig:res_scaled_values}) are sufficiently large to explain observed eccentricities, and we highlight that this did not have to be the case---the unnormalized free eccentricities (Fig.\:\ref{fig:data}) are $\sim 0.01$, and in principle could have been significantly higher.

Nevertheless, observed free eccentricities generically lie below the prediction for adiabatic MMR crossings in Fig.\:\ref{fig:res_scaled_values}.
Because the relevant timescale for adiabaticity is $\sim 10^5$ orbits for Earth-mass planets (Eq.\:\ref{eq:tacrit}), it may be that migration is not adiabatic, leading to smaller eccentricities (Fig.\:\ref{fig:timescale}).
Additionally, most migration mechanisms (e.g., tides, disk-planet interactions) also drive eccentricity damping.
It may therefore be that the eccentricities are explained through the combination of a kick as the MMR is crossed, followed by residual damping of the eccentricity (as opposed to $\sim 100$ eccentricity damping timescales that would erase any kick), see Fig.\:\ref{fig:edamping} and Sec.\:\ref{sec:edamping}. 



Our proposed scenario of jumps across MMRs is in some ways similar to that of \cite{Petrovich13}.
In that work, the authors instead considered the growth of MMRs as the planets accreted mass in their early history.
In that picture, the eccentricities also get kicked to higher values, because the edge of the MMR eventually reaches the planets' initial $\Delta$, leading to the same scaling of $\mathcal{Z}_\textrm{Jump}$ with mass as we obtain above.
However, such a model would presumably predict that we should find most planet pairs in the pileup close to the edge (separatrix) of the MMR that they jumped over.
In our picture the migration needs to be sufficient to reach the MMR, but it has no reason to halt once it reaches it, and should lead to a broader distribution of $\Delta$ values.
We see in Fig.\:\ref{fig:res_scaled_values} that indeed, observed values of $\Delta/\Delta_\textrm{Jump}$ span a broad range, rather than being clustered at $\approx 1$.

\hfill \break

This work was funded in part by The Rose Hills Foundation Science and Engineering Summer Undergraduate Research Fellowship program.
The presented numerical calculations were made possible by computational resources provided through an endowment by the Albrecht family.
This research has made use of the NASA Exoplanet Archive, which is operated by the California Institute of Technology, under contract with the National Aeronautics and Space Administration under the Exoplanet Exploration Program.

\bibliography{Bib,references}
\bibliographystyle{aasjournal}

\appendix
\section{MMR Dynamics}

In this Appendix we obtain analytical expressions for the size of the jumps in eccentricity and $\Delta$ when crossing a first-order MMR, as well as for the equilibrium eccentricity expected from strong eccentricity damping.

To lowest order in eccentricity, the dynamics of a first-order $j$:$j-1$ MMR between a pair of planets of mass $m_1$ and $m_2$ around a star of mass $M_*$ can be modeled as a one degree-of-freedom system \citep{henrard_second_1983}.
In the notation of \citet{hadden_integrable_2019}, who also generalizes such models to higher-order MMRs, the Hamiltonian is given by

\begin{equation}
    H(J,\theta) = -\frac{1}{2}A(J-J^*)^2 - \tilde{\epsilon}\sqrt{J}\cos\theta
    \label{eq:ham_fo_res}~
\end{equation}
where the resonant angle $\theta$ is the angle at which conjunctions occur, and $J = \eta^2 |\mathcal{Z}|^2$ is the conjugate momentum variable, with \citep{Hadden19}
\begin{equation}
\eta^2 \equiv \frac{f^2 + g^2}{\frac{f^2}{\beta_1}\Big(\frac{j}{j-1}\Big)^{1/3}  + \frac{g^2}{\beta_2}},
\end{equation}
$\beta_i = m_i/(m_1+m_2)$, and the $f$ and $g$ coefficients are the same as above (and depend on the particular resonance). 
The parameters $A$ and $\tilde\epsilon$ appearing in Equation \eqref{eq:ham_fo_res} are related to planet masses and the particular MMR according to 

\begin{eqnarray}
   A&=&\frac{3j}{2}\left[\frac{j}{\beta_2} +\frac{j-1}{\beta_1}\Big(\frac{j}{j-1}\Big)^{1/3} \right] 
   \nonumber \\
   \tilde{\epsilon}&=&2\left[\frac{f^2}{\beta_1}\Big(\frac{j}{j-1}\Big)^{1/3} + \frac{g^2}{\beta_2}\right]^{1/2}\frac{m_1m_2}{M_*(m_1+m_2)},
\end{eqnarray}
and finally, the ``proximity" to resonance $J^\star$ is a conserved quantity relating $\mathcal{Z}$ and $\Delta$:
\begin{equation}
J^\star = \eta^2 |\mathcal{Z}|^2 - \frac{j}{A}\Delta
\end{equation}

Hamilton's equations remain valid if the Hamiltonian function and momentum variables are re-scaled by a common factor.  This fact allows the dynamics of any first-order MMR to be described by the common model Hamiltonian 
\begin{equation}
\hat{H} = -\frac{1}{2}(\hat{J}-\hat{J}^*)^2 -\sqrt{\hat{J}}\cos\theta,
\end{equation}  
where $\hat{J} = (A/\tilde{\epsilon})^{2/3}J$, $\hat{J}^* = (A/\tilde{\epsilon})^{2/3} J^*$, provided we define a dimensionless ``time" variable $\tau = t/T_\textrm{res}$ \citep{henrard_second_1983}, where the resonant timescale is given by

\begin{equation}
T_\textrm{res} \equiv A^{-1/3}\tilde{\epsilon}^{-2/3}. \label{eq:Tres}
\end{equation}

The equilibrium near $\mathcal{Z}=0$ disappears when $\hat{J}^* = 3/2^{4/3}$ \citep{Ferraz07}. For this value of $\hat{J}^* = 3/2^{4/3}$, we find using the expressions in Appendix B of \cite{lammers_instability_2024} that $\hat{J}$ reaches a maximum on the lower branch of the separatrix, corresponding to a maximum value of $\mathcal{Z}_\textrm{Jump}$ of
\begin{eqnarray}
\mathcal{Z}_\textrm{Jump} &\approx& \frac{2.67}{\eta} \Bigg ( \frac{\tilde{\epsilon}}{A} \Bigg)^{1/3}. \label{eq:zjump}
\end{eqnarray}
Given the conservation of $\hat{J}^* = 3/2^{4/3}$ (in the adiabatic limit where $\hat{J}^*$ changes slowly), this corresponds to a maximum value of $\Delta_\textrm{Jump}$ of
\begin{eqnarray}
\Delta_\textrm{Jump} &\approx& \frac{2.38}{j}A^{1/3} \tilde{\epsilon}^{2/3}. \label{eq:deltajump}
\end{eqnarray}
The equilibrium eccentricity to which strong eccentricity damping would drive planet pairs is given by
\begin{equation}
\mathcal{Z}_{eq} = \frac{\tilde{\epsilon}}{2j\eta}\frac{1}{\Delta}
\end{equation}

\end{document}